\documentclass[a4paper,12pt]{article}

\usepackage{amsmath,amssymb,mathtools} 
\usepackage{color}
\usepackage{graphicx}
\usepackage{subfigure}
\usepackage{cite}           
\usepackage{hyperref}            
\usepackage{multirow,makecell}   
\usepackage{textcomp}
\usepackage{wasysym}
\usepackage{setspace}
\usepackage{verbatim}
\usepackage{float}
\usepackage{ulem}
\usepackage[utf8]{inputenc}
\graphicspath{{newFig/}}

\usepackage[text={17.2cm,25.2cm},centering]{geometry}

\numberwithin{equation}{section}

\begin{document}
	\title{\textbf{Drag force and heavy quark potential in a rotating background}}
	\author{Jun-Xia Chen$^{1}$\footnote{chenjunxia@mails.ccnu.edu.cn }, De-Fu Hou$^{1}$\footnote{houdf@mail.ccnu.edu.cn }, Hai-Cang Ren$^{1,2}$\footnote{renhc@mail.ccnu.edu.cn } }
	\date{}
	
	\maketitle
	
	\vspace{-10mm}
	
	\begin{center}
	{\it
		$^{1}$ Institute of Particle Physics and Key Laboratory of Quark and Lepton Physics (MOS),
		Central China Normal University, Wuhan 430079, China\\ \vspace{1mm}
		
		$^{2}$ Physics Department, The Rockefeller University,
		1230 York Avenue, New York, NY 10021-6399\\ \vspace{1mm}			
	}
	\vspace{10mm}
\end{center}

\begin{abstract}
	
	We explored the gravity dual of a rotating quark-gluon plasma by transforming the boundary coordinates of the large black hole limit of Schwarchild-$\text{AdS}_5$ metric. The Euler-Lagrange equation of the Nambu-Goto action and its solution become more complex than those without rotation. For small angular velocity, we obtained an analytical form of the drag force acting on a quark moving in the direction of the rotation axis and found it stronger than that without rotation. We also calculated the heavy quark potential under the same approximation. For the quarkonium symmetric with respect to the rotation axis, the depth of the potential is reduced by the rotation. For the quarkonium oriented in parallel to the rotation axis, the binding force is weakened and the force range becomes longer. We also compared our holographic formulation with others in the literature.
	
\end{abstract}

\baselineskip 18pt
\thispagestyle{empty}
\newpage

\tableofcontents

\maketitle

\section{Introduction}\label{sec:01_intro}

The relativistic heavy ion collision has been an important experimental probe to the non-perturbative properties of QCD under unusual conditions. Rotation is one of them as the off-central collisions tend to deposit high angular momentum in the hot and dense QCD matter left behind \cite{STAR:2017ckg,Liang:2004ph}. A theoretical description of the physics of such a system including the axial-vortical effect and polarization effect remains highly challenging. See \cite{Becattini:2021lfq} and the references therein.

Let $\rho$ be the density operator of a thermal ensemble with a rotational invariant Hamiltonian $H$. In the Schroedinger representation, we have the Louville equation
\begin{equation}
	\frac{\partial\rho}{\partial t}+i[H,\rho]=0.
\end{equation}
Under a rotation with the angular velocity $\omega$ about $z$-axis, $\rho\to e^{i\omega tJ_z}\rho e^{-i\omega tJ_z}$, we have
\begin{equation}
	\frac{\partial\rho}{\partial t}+i[H-\omega J_z,\rho]=0,
\end{equation}
with density operator at the thermal equilibrium
\begin{equation}
	\rho = e^{-\beta(H-\omega J_z)}.
	\label{ensemble}
\end{equation}
It can be shown explicitly that for the canonical form (dictated by Noether theorem) of the angular momentum operator $J_z$, the path-integral form of the ensemble average defined by $\rho$ corresponds to the Lagrangian in a non-relativistic rotating frame given by
\begin{equation}
	x_i\to M_{ij}x_j
	\label{rot_NR1},
\end{equation}
with the $3\times 3$ matrix
\begin{equation}
	M=\left(\begin{array}{ccc}
		\cos\omega t & -\sin\omega t & 0\\
		\sin\omega t & \cos\omega t  & 0\\
		0 & 0 & 1
	\end{array}\right),
	\label{rot_NR2}
\end{equation}
which transforms the spacetime metric to
\begin{equation}
	ds^2=-[1-\omega^2(x^2+y^2)]dt^2+2\omega(xdy-ydx)dt+dx^2+dy^2+dz^2,
	\label{metric_NR}
\end{equation}
with $\omega^2(x^2+y^2)<1$. The metric Eq.(\ref{metric_NR}) has been widely employed in the Lattice QCD \cite {Braguta:2021jgn,Braguta:2023yjn} and the mean field theory of NJL model \cite {Jiang:2016wvv,Zhang:2018ome} as well as other field theoretic investigations \cite{ Rybalka:2018uzh,Fujimoto:2021xix,Chen:2020xsr,Chen:2023cjt,Yang:2023zqe,Buzzegoli:2022omv,BitaghsirFadafan:2008adl, Hou:2021own,Chernodub:2020qah,Giantsos:2022qgq}. The adjective "non-relativistic" is not a concern for our purpose since it merely serves a coordinate transformation to implement the path integral formulation of the thermal ensemble Eq.(\ref{ensemble}) and can be used beyond the non-relativistic limit.

In this work, we shall explore the drag force and heavy quark potential in a rotating quark-gluon plasma (QGP) by means of the AdS/CFT duality \cite{Maldacena:1997re,Witten:1998zw}. We shall use the $N=4$ supersymmetric Yang-Mills theory as a proxy of QCD in the deconfined phase. In the absence of rotation, the large $N_c$ super Yang-Mills in strong 't Hooft coupling at a nonzero temperature corresponds to the background 5D metrics of a non-extremal 3D black brane, i.e.
\begin{equation}
	ds^2=\frac{r^2}{R^2}\left(-f(r)dt^2+d\vec x^2\right)+\frac {1}{f(r)}\frac{R^2}{r^2}dr^2,
	\label{blackbrane}
\end{equation}
with
\begin{equation}
	f(r)=1-\frac{r_t^4}{r^4},
	\label{blackening function}
\end{equation}
where $R$ is the AdS radius and $r_t$ is the position of the black brane horizon. This background stems from a low energy solution of type-IIB superstring in 10 dimensions \cite{Horowitz:1991cd} with the additional 5 dimensions taking the form $S^5$. The 5D metric Eq.(\ref{blackbrane}) also follows from the large black hole limit of the standard Schwarzschild-$\text{AdS}_5$ metric \cite{Witten:1998zw}. The physics of the $N=4$ super Yang-Mills resides on the conformal boundary $r\to\infty$, which is planar, $R^1\times R^3$, and the $S^5$ sector of the solution corresponds to its R-charge. While there are spinning brane solutions from the superstring in the literature \cite{Gubser:1998jb,Brandhuber:1999jr,Duff:1999rk}, the rotation there is in the sector $S^5$ describing a nonzero R-charge density of the super Yang-Mills, and thereby is not a spatial rotation. From bottom-up perspective, there has not been a consensus way to introduce spatial rotation into holographic formulations. One approach in the literature is to make a local Lorentz transformation to the static frame of a small segment of the rotating QGP \cite{BravoGaete:2017dso,Erices:2017izj,Awad:2002cz,Zhao:2022uxc,Chen:2020ath,Chen:2022mhf,Zhou:2021sdy,Zhang:2023psy}. Another approach is to replace the black brane metric Eq.(\ref{blackbrane}) with the Kerr-$\text{AdS}_5$ metric \cite{Hawking:1998kw,Arefeva:2020jvo,Golubtsova:2022ldm,Golubtsova:2021agl}. Both formulations suffer from tangible limitations as we shall see. In what follows, we shall introduce a rotation by simply transforming the boundary coordinates of the metric Eq.(\ref{blackbrane}) according to Eqs.(\ref{rot_NR1}) and (\ref{rot_NR2}), and the transformed metric will be referred to as the global rotation background with the boundary metric exactly Eq.(\ref{metric_NR}). So the boundary physics is that of the $N=4$ super Yang-Mills in a flat spacetime formulated in a rotating frame of reference. This metric is closely related to the thermal ensemble with a nonzero angular momentum in view of the above discussion and the detailed analysis in the next section. 

The drag force and the heavy quark potential are extracted from the solutions of the Euler-Lagrange equations of the Nambu-Goto action of a string world-sheet embedded in the 5d metric Eq.(\ref{blackbrane}). A technical difficulty arises because of the inhomogeneity introduced by the rotation. The simplistic solution ansatz in the absence of rotation does not work anymore and one has to solve the full set of the Euler-Lagrange equations and the analytical tractability is limited to small $\omega$ expansion. For a heavy quark moving in the direction of the rotation axis, we derived the analytical formulas of the drag force up to the order $\omega^2$ with the $\omega^2$ correction suggesting a higher rate of energy loss because of rotation. In addition, the centrifugal force also emerges in our solution as expected. For the heavy quark potential, we derived the explicit integral representations of the $O(\omega^2)$ correction by the rotation. Both reduction and enhancement by rotation are observed depending on the location and orientation of the heavy quarkonium.

This paper is organized as follows. An overview of the global rotation background is provided in the next section. The drag force and the heavy quark potential are calculated in sections \ref{sec:03} and \ref{sec:04}. Other rotation formulations will be commented in section \ref{sec:05} and section \ref{sec:06} conclude the paper. Some technical details behind the heavy quark potential are deferred to Appendices A and B.    

\section{Global rotation frame}
\label{sec2}

\subsection{A grand canonical ensemble with a nonzero angular momentum} 

In this subsection, we prove the equivalence between the canonical form of the partition function of the thermal ensemble with a macroscopically nonzero angular momentum $J_z$,
\begin{equation}
	\mathcal{Z}=\rm{Tr}e^{-\frac{H-\omega J_z}{T}},
	\label{partition}
\end{equation}
and the path integral form in the co-rotation frame defined by the transformation Eq.(\ref{rot_NR2}), and consequently by the metric Eq.(\ref{metric_NR}). 

Consider a rotation invariant Lagrangian density with field components $\Phi_a$ ($a=1,2,...$), bosonic or fermionic,
\begin{equation}
	\mathcal{L}=\mathcal{L}(\dot{\Phi}(X),\Phi(X)),
	\label{lagrangian}
\end{equation}
where $\Phi$, $X=(t,\vec R)$ refer to the static frame of reference. The canonical momentum
\begin{equation}
	\Pi_a=\frac{\partial\mathcal{L}}{\partial\dot{\Phi}_a},
	\label{momentum}
\end{equation}
and the Hamiltonian density follows from the Legendre transformation 
\begin{equation}
	\mathcal{H}=\Pi_a\dot{\Phi}_a-\mathcal{L}=\tilde{\Pi}\dot{\Phi}-\mathcal{L}=\mathcal{H}(\Pi,\Phi),
	\label{hamiltonian}
\end{equation} 
with
\begin{equation}
	\dot{\Phi}_a=\mathcal{F}_a(\Pi,\Phi),
	\label{velocity}
\end{equation} 
where $\mathcal{F}_a(\Pi,\Phi)$ is obtained by inverting the functional relationship between $\Pi$ and $\dot{\Phi}$ specified by Eq.(\ref{momentum}). Here and in what follows, the notation of field or its canonical momentum without a subscript refers to a column matrix made of all components with a tilde signifying its transpose. The spatial integration of Eq.(\ref{hamiltonian}) gives rise to the Hamiltonian $H(\Pi,\Phi)$ in the static frame.

In terms of the field $\phi$ and spacetime coordinates $x=(t,\vec r)$ of the rotating frame, we have
\begin{equation}
	\Phi(t,\vec R)=e^{-iS_z\omega t}\phi(t,\vec r),
	\label{Phi_rot}
\end{equation}
where $X$ and $x$ are related via the rotation Eq.(\ref{rot_NR2}), and $S_z$ is the spin matrix. It is straightforward to show that
\begin{equation}
	\left(\frac{\partial\Phi}{\partial t}\right)_{\vec R}
	=e^{-iS_z\omega t}\left(\left(\frac{\partial\phi}{\partial t}\right)_{\vec r}-i\omega\mathcal{J}_z\phi\right),
	\label{Phi_dot}
\end{equation}
with the angular momentum operator
\begin{equation}
	\mathcal{J}_z=-i\left(x\frac{\partial}{\partial y}-y\frac{\partial}{\partial x}\right)+S_z.
\end{equation}
Substituting Eq.(\ref{Phi_rot}) and Eq.(\ref{Phi_dot}) into Eq.(\ref{lagrangian}), the Lagrangian density becomes
\begin{equation}
	\mathcal{L}=\mathcal{L}\left[e^{-iS_z\omega t}\left(\dot{\phi}-i\omega \mathcal{J}_z\phi\right),e^{-iS_z\omega t}\phi\right]=\mathcal{L}(\dot{\phi}-i\omega \mathcal{J}_z\phi,\phi),
	\label{lagrangian_rot}
\end{equation}
where the last step follows from the rotation invariance.
The canonical momentum and Hamiltonian density in the rotating frame are
\begin{equation}
	\Pi^{(\rm{rot})}_a=\frac{\partial\mathcal{L}}{\partial\dot{\phi}_a},
\end{equation}
and
\begin{equation}
	\mathcal{H}^{(\rm{rot})}=\tilde{\Pi}^{(\rm{rot})}\dot{\phi}-\mathcal{L}.
	\label{legendre}
\end{equation}
It follows from Eq.(\ref{momentum}), Eq.(\ref{velocity}) and Eq.(\ref{lagrangian_rot}) that
\begin{equation}
	(\dot{\phi}-i\omega \mathcal{J}_z\phi)_a =\mathcal{F}_a(\Pi^{(\rm{rot})},\phi),
\end{equation}
and the Legendre transformation Eq.(\ref{legendre}) yields
\begin{equation}
	\mathcal{H}^{(\rm{rot})}=\mathcal{H}\left(\Pi^{(\rm{rot})},\phi\right)+i\omega \tilde{\Pi}^{(\rm{rot})}\mathcal{J}_z\phi.
	\label{hamiltonian_rot}
\end{equation}
The first term of Eq.(\ref{hamiltonian_rot}) depends on $\Pi^{(\rm{rot})}$ and $\phi$ the same way as the Hamiltonian density in the static frame on $\Pi$ and $\Phi$, and the second term is just $-\omega$ times the field theoretic angular momentum density following Noether theorem. The spatial integration of Eq.(\ref{hamiltonian_rot}) gives rise to
\begin{equation}
	H^{(\rm{rot})}=H\left(\Pi^{(\rm{rot})},\phi\right)-\omega J_z.
\end{equation}

For quantized field with $H=H^{(\rm{rot})}$, the transformation from the canonical formulation of Eq.(\ref{partition}) to the path integral amounts to undo the Legendre transformation Eq.(\ref{legendre}) and one ends up with the Lagrangian density Eq.(\ref{lagrangian_rot}) post the "non-relativistic" rotation. The superscript "rot" is immaterial since $\Pi^{(\rm{rot})}$ and $\phi$ satisfy the same canonical commutation relation as $\Pi$ and $\Phi$ do. The equivalence claimed at the beginning of this section is thereby proved.

Following the equivalence, the coordinates $\vec r$ pertain to the laboratory frame where a QGP fireball is in rotation and the field operators $\phi$ and $\Pi^{(\rm rot)}$ are defined there. The mathematical form of the Lagrange underlying the path integral is defined in a non-relativistic rotating frame according to Eqs.(\ref{rot_NR1}) and (\ref{rot_NR2}). However, its usage is not limited to non-relativistic physics.

For QCD, $\Phi$ in Eq.(\ref{lagrangian}) consists of gluon field $A_\mu$ and quark fields ($\psi$, $\bar\psi$). For a free Dirac field with $\Phi=(\psi,\bar\psi)$, it can be shown explicitly that the Dirac Lagrangian density transformed by Eqs.(\ref{rot_NR1}), (\ref{rot_NR2}) and (\ref{Phi_rot}) is identical to the one derived in the literature from the vielbein formulation of the metric Eq.(\ref{metric_NR}). 

\subsection{Background geometry and Nambu-Goto action}\label{sec:02}

For a holographic description of a QGP carrying a macroscopic angular momentum, we transform the black brane metric Eq.(\ref{blackbrane}) into a uniformly rotating frame with respect to boundary coordinates. Starting with Eq.(\ref{blackbrane}) in terms of the cylindrical coordinates of $\vec x$
\begin{equation}
	\label{eq22}
	ds^{2}=-\frac{r^{2}}{R^{2}}f(r)dt^{2}+\frac{r^{2}}{R^{2}}(dl^{2}+l^{2}d\phi^{2}+dz^{2})+\frac{1}{f(r)}\frac{R^{2}}{r^{2}}dr^{2},
\end{equation}	
with $\vec x=(l\cos\phi,l\sin\phi,z)$, the transformation (\ref{rot_NR2}) amounts to
\begin{equation}
	(t,l,\phi,z,r)\to (t,l,\phi+\omega t,z,r).
\end{equation}
The resulting metric reads
\begin{eqnarray}
	\label{eq222}
	ds^{2} &=& \frac{r^{2}}{R^{2}}[-(f(r)-\omega^{2}l^{2})dt^{2}+l^{2}d\phi^{2}+2\omega l^{2} dt d\phi+dl^{2}+dz^{2}]+\frac{1}{f(r)}\frac{R^{2}}{r^{2}}dr^{2}\nonumber
	\\&=& \frac{r^2}{R^2}\{ -[f(r)-\omega^2(x^2+y^2)]dt^2+2\omega dt(xdy-ydx)+dx^2+dy^2+dz^2\}\nonumber\\
	&+&
	\frac{1}{f(r)}\frac{R^{2}}{r^{2}}dr^{2},
\end{eqnarray} 
where we have also restored the cartesian form of the rotating metric in the second and third lines for later usage. The Hawking temperature of the black hole
\begin{equation}
	\label{eq223}
	T=\frac{r_{t}}{\pi R^{2}},
\end{equation} 
where $r_t$ is the position of the black brane horizon in Eq.(\ref{blackening function}). 

The Nambu-Goto action underlying the drag force and heavy quark potential is the area of the world-sheet embedded in the spacetime metric Eq.(\ref{eq222}), given by
\begin{equation}
	S=-\frac{1}{2\pi\alpha^\prime}\int dtdr\sqrt{-g},
	\label{NambuGoto}
\end{equation}
where $g$ is the determinant of the world-sheet metric, and $\alpha^\prime$ is related to the 't Hooft coupling constant
\begin{equation}
	\label{lambda}
	\lambda=\frac{R^4}{\alpha^{\prime 2}}.
\end{equation}
The world-sheet satisfying the Euler-Lagrange equations of $S$ gives rise to the drag force and heavy quark potential of the $N=4$ super Yang-Mills at large $N_c$ and large $\lambda$. Taking $(t,r)$  as world-sheet coordinates, and $(z, \phi, l)$ as functions of them, the world-sheet metric takes the form 
\begin{eqnarray}
	ds^2 &=& \frac{r^2}{R^2}\Big[(\omega^2l^2-f+2\omega l^2\dot{\phi}+l^2\dot{\phi}^2+\dot{l}^2+\dot{z}^2)dt^2\nonumber\\
	&+&\left(l^2\phi^{\prime 2}+l^{\prime 2}+z^{\prime 2}+\frac{R^4}{r^4f}\right)dr^2+2(\omega l^2\phi^\prime+l^2\dot{\phi}\phi^{\prime}+\dot{l}l^\prime+\dot{z}z^\prime)dtdr\Big],
\end{eqnarray}
and
\begin{eqnarray}
	g &=& \frac{r^4}{R^4}\Big[(\omega^2l^2-f+2\omega l^2\dot{\phi}+l^2\dot{\phi}^2+\dot{l}^2+\dot{z}^2)\left(l^2\phi^{\prime 2}+l^{\prime 2}+z^{\prime 2}+\frac{R^4}{r^4f}\right)\nonumber\\&
	-&(\omega l^2\phi^\prime+l^2\dot{\phi}\phi^{\prime}+\dot{l}l^\prime+\dot{z}z^\prime)^2\Big],
	\label{NGdet}
\end{eqnarray}
where the dot and prime denote the derivatives with respect to $t$ and $r$ respectively. As the determinant Eq.(\ref{NGdet}) is independent of $z$ and $\phi$, the Euler-Lagrange equations of motion read
\begin{equation}
	\frac{d}{dt}\left(\frac{\partial\sqrt{-g}}{\partial\dot{z}}\right)+\frac{d}{dr}\left(\frac{\partial\sqrt{-g}}{\partial z^\prime}\right)=0,
	\label{EL1}
\end{equation}
\begin{equation}
	\frac{d}{dt}\left(\frac{\partial\sqrt{-g}}{\partial\dot{\phi}}\right)+\frac{d}{dr}\left(\frac{\partial\sqrt{-g}}{\partial \phi^\prime}\right)=0,
	\label{EL2}
\end{equation}
and
\begin{equation}
	\frac{d}{dt}\left(\frac{\partial\sqrt{-g}}{\partial\dot{l}}\right)+\frac{d}{dr}\left(\frac{\partial\sqrt{-g}}{\partial l^\prime}\right)-\frac{\partial\sqrt{-g}}{\partial l}=0.
	\label{EL3}
\end{equation}
One complication arises from the metric Eq.(\ref{eq222}) is the coupling of the three Euler-Lagrange equations (\ref{EL1}), (\ref{EL2}) and (\ref{EL3}) at a nonzero angular velocity $\omega$, which hinders the utilization of the simplistic solution ansatz employed in the absence of rotation \footnote{The same complication was recognized and taken into account in the Kerr-$\text{AdS}_5$ formulation discussed in \cite{Arefeva:2020jvo}.}. To demonstrate this point, we assume that
\begin{equation}
	\phi=l={\rm{const.}}, \qquad z=vt+\zeta(r),
	\label{ansatz}
\end{equation}
with constant $v$. This ansatz covers three types of world-sheet in the absence of rotation: 1) the drag force acting on a heavy quark moving in z-direction at constant speed $v$; 2) a pair of static heavy quark and antiquark separated in z-direction ($v=0$) and 3) a static heavy quark or antiquark ($v=0$ and $z=\rm{const.}$). It follows from the conditions $\dot{\zeta}=\dot{\phi}=\dot{l}=0$, that the determinant $g$ as well as $\frac{\partial g}{\partial\dot{z}}$, $\frac{\partial g}{\partial\dot{\phi}}$ and $\frac{\partial g}{\partial\dot{l}}$  are time independent so the first terms of Eqs.(\ref{EL1}), (\ref{EL2}) and (\ref{EL3}) remain vanishing in the presence of rotation. On the other hand, the conditions $\phi^\prime=l^\prime=0$ Eq.(\ref{ansatz}) of the ansatz imply that   
\begin{equation}
	\frac{\partial g}{\partial\phi^\prime}=-\frac{2r^4}{R^4}\omega l^2v\zeta^\prime,
\end{equation}
and
\begin{equation}
	\frac{\partial g}{\partial l^\prime}=0, \qquad 
	\frac{\partial g}{\partial l}=2\omega^2\frac{r^4}{R^4}\left( \zeta^{\prime 2}+\frac{R^4}{r^4f}\right)l.
\end{equation}
Consequently, the simplistic ansatz Eq.(\ref{ansatz}) with $\zeta^\prime$ determined by Eq.(\ref{EL1}) fails to satisfy Eq.(\ref{EL2}) and Eq.(\ref{EL3}) for $v\neq 0$ and fails to satisfy Eq.(\ref{EL3}) for $v=0$, unless $\omega=0$ or $l=0$.

In what follows, we shall make small $\omega$ approximation. As shown in Appendix A, for the world-sheet not extending to the horizon, the simplistic solution in the absence of rotation suffices to evaluate the Nambu-Goto action to the order $\omega^2$. This is the case for the "U-shaped world-sheet" underlying the heavy quark potential. For drag force and the self energy of a single heavy quark (used for subtracting the UV divergence of the heavy quark potential), the world-sheet extends to the horizon, a perturbative solution of Eqs.(\ref{EL1})-(\ref{EL3}) up to order $\omega^2$ will be found in the subsequent sections.

\section{Drag force in global rotating background}
\label{sec:03}

When a heavy quark traverses a hot QGP, its momentum and energy are dissipated through friction with the medium. Such a friction is quantified by a drag force. From holographic perspective, the gravity dual of the heavy quark trajectory is a world-sheet extending from the trajectory on the AdS boundary to the black hole horizon. Thereby the drag force corresponds to the canonical momentum density of the Nambu-Goto action underlying the world-sheet \cite{Arefeva:2020jvo}, i.e.
\begin{equation}
	\label{eq60}
	\mathfrak{f}_\mu=-\frac{1}{2\pi\alpha^\prime }\frac{\partial \mathcal{L}}{\partial (\frac {\partial X^{\mu}}{\partial r} )},
\end{equation}
where $\mathcal{L}\equiv\sqrt{-g}$ with $g$ the determinant of the induced metric Eq.(\ref{NGdet}) and the derivatives are evaluated at the solution of the Euler-Lagrange equation. 

We consider a heavy quark moving in $z$ direction with speed $v$, taking the generic solution ansatz,
\begin{equation}
	\label{eq601}
	z=vt+\xi (r), \quad l=l(r), \quad \phi =\phi (r),
\end{equation}
the determinant of the induced metric is 
\begin{equation}
	\label{eq602}
	g=-\frac{r^{4}}{R^{4}}\left[(f-\omega^{2}l^{2}-v^{2})\left(l^{2}\phi^{\prime2}+l^{\prime2}+\xi^{\prime2}+\frac{R^{4}}{r^{4}f}\right)+(\omega l^{2}\phi^{\prime}+v\xi^{\prime})^{2}\right].
\end{equation}
It follows from Eq.(\ref{eq60}) with $X^\mu=z, \phi, l$ of the solution ansatz Eq.(\ref{eq601}) that the components of the drag force to be evaluated read
\begin{equation}
	\begin{split}
		\label{eq6021}
		&\mathfrak{f}_z = -\frac{1}{2\pi\alpha^\prime}\frac{\partial\mathcal{L}}{\partial z^\prime},\\&
		\mathfrak{f}_\phi = -\frac{1}{2\pi\alpha^\prime}\frac{\partial\mathcal{L}}{\partial \phi^\prime},\\&
		\mathfrak{f}_l = -\frac{1}{2\pi\alpha^\prime}\frac{\partial\mathcal{L}}{\partial l^\prime},
	\end{split}
\end{equation}
where $\mathfrak{f}_\phi$, being a generalized force with respect to the angle, represents actually a torque.

Because of the coupling among different components, $\xi$, $\phi$ and $l$, discussed in the preceding section, one has to solve the set of three highly nonlinear ordinary differential equations (\ref{EL1})-(\ref{EL3}). Instead, we focus on the small $\omega$ approximation for analytical tractability. 

If $\phi\equiv\omega\chi$ is substituted, the determinant Eq.(\ref{eq602}) becomes a function of $\omega^2$, $\xi^\prime$, $\chi^\prime$, $l^\prime$ and $l$ and we expect series expansion of the solution for $\xi^\prime$, $\chi^\prime$, $l^\prime$ and $l$ in $\omega$ contains only even powers. Consequently, the solution ansatz up to the leading order correction by rotation reads 	
\begin{equation}
	\begin{split}
		\label{eq603}
		& z=vt+\xi_{0}(r)+\omega^{2}\xi_{1}(r),
		\\& \phi=\phi_0+\omega \phi_1(r),
		\\& l=l_0+\omega^2 l_1(r),
	\end{split}
\end{equation}
where $l_0$ and $\phi_0$ are constants, and $\xi_{0}(r)$ is the solution \cite{Gubser:2006bz,Herzog:2006gh} in the absence of rotation,
\begin{equation}
	\label{eq6031}
	\xi_{0}^{\prime}=v\frac{R^{2}r_{t}^{2}}{r^{4}-r_{t}^{4}}.
\end{equation}
Substituting Eq.(\ref{eq603}) into Eq.(\ref{eq602})
\begin{eqnarray}
	g&=&-\frac{r^{4}}{R^{4}}\bigg\{[f-\omega^{2}(l_{0}+\omega^{2}l_{1})^{2}-v^{2}]\left[(l_{0}+\omega^{2}l_{1})^{2}\omega^{2}\phi_{1}^{\prime2}+\omega^{4}l_{1}^{\prime2}+(\xi_{0}^{\prime}+\omega^{2}\xi_{1}^{\prime})^{2}+\frac{R^{4}}{r^{4}f}\right]\nonumber\\
	& +&[\omega^{2}(l_{0}+\omega^{2}l_{1})^{2}\phi_{1}^{\prime}+v(\xi_{0}^{\prime}+\omega^{2}\xi_{1}^{\prime})]^{2}\bigg\}.
	\label{det_1}
\end{eqnarray}
The determinant $g$ does not contain $\phi_1$ and $\xi_1$, so we have 
\begin{equation}
	\label{first_integral}
	\frac{\partial \mathcal{L}}{\partial\phi_{1}^{\prime}}  =\rm{const.}, \quad \frac{\partial \mathcal{L}}{\partial\xi_{1}^{\prime}}  =\rm{const.},
\end{equation}
and the Euler-Lagrange equation for $l_1$ reads
\begin{equation}
	\frac{d}{dr}(\frac{\partial\mathcal{L}}{\partial l_1^\prime})-\frac{\partial\mathcal{L}}{\partial l_1}=0.
	\label{EL4}
\end{equation}
In what follows we shall determine the explicit forms of $\phi_1^\prime$, $ \xi_1^\prime$  and $l_1^\prime$.

Expanding the determinant Eq.(\ref{det_1}) to the order $\omega^4$, i.e.
\begin{equation}
	g=g_0+\omega^2g_1+\omega^4 g_2,
\end{equation}
we find that
\begin{equation}
	\begin{split}
		&g_0=-1+v^2,
		\\&g_1=\frac{r^4}{R^4}\left[l_0^2(v^2-f)\phi_1^{\prime 2}-2vl_0^2\xi_0^\prime\phi_1^\prime-2f\xi_0^\prime\xi_1^\prime+l_0^2\left(\xi_0^{\prime 2}+\frac{R^4}{r^4f}\right)\right],
		\\&g_2 = \frac{r^4}{R^4}\bigg[(v^2-f)(2l_0l_1\phi_1^{\prime 2}+l_1^{\prime 2}+\xi_1^{\prime 2})+2\left(\xi_0^{\prime 2}+\frac{R^4}{r^4f}\right)l_0l_1
		\\&\quad +l_0^2(l_0^2\phi_1^{\prime2}+2\xi_0^\prime\xi_1^\prime)-4vl_0\xi_0^\prime l_1\phi_1^\prime-(v\xi_1^\prime+l_0^2\phi_1^\prime)^2\bigg].
	\end{split}
\end{equation}
Correspondingly
\begin{equation}
	\mathcal{L}=\sqrt{-g}=\mathcal{L}_0+\omega^2\mathcal{L}_1+\omega^4\mathcal{L}_2,
	\label{L_expansion}
\end{equation}
where
\begin{equation}
	\begin{split}
		&\mathcal{L}_0=\sqrt{-g_0}=\sqrt{1-v^2},
		\\&\mathcal{L}_1=-\frac{g_1}{2\sqrt{1-v^2}},
		\\&\mathcal{L}_2=-\frac{1}{2\sqrt{1-v^2}}\left[g_2+\frac{g_1^2}{4(1-v^2)}\right].
	\end{split}
	\label{L_expansion1}
\end{equation}
The drag force components Eq.(\ref{eq6021}) in accordance with Eqs.(\ref{eq603}) and (\ref{L_expansion}) take the form
\begin{equation}
	\begin{split}
		\mathfrak{f}_z &= -\frac{1}{2\pi\alpha^\prime}\left(\frac{\partial\mathcal{L}_1}{\partial \xi_1^\prime}+\omega^2\frac{\partial\mathcal{L}_2}{\partial \xi_1^\prime}\right),\\
		\mathfrak{f}_\phi &= -\frac{1}{2\pi\alpha^\prime}\omega\frac{\partial\mathcal{L}_1}{\partial \phi_1^\prime},\\
		\mathfrak{f}_l &= -\frac{1}{2\pi\alpha^\prime}\omega ^2\frac{\partial\mathcal{L}_2}{\partial l_1^\prime},
		\label{eq6022}
	\end{split}
\end{equation}
where the absence of $\mathcal{L}_0$ follows from its constancy and the absence of $\mathcal{L}_1$ in the last equation follows from the observation that $l_1^\prime$ shows up only in $\mathcal{L}_2$ through $g_2$. Only the leading order terms in $\omega$ are maintained in Eqs.(\ref{eq6022}).

The first equation of Eqs.(\ref{first_integral}) to the order $\omega^2$ of Eq.(\ref{L_expansion}) gives rise to
\begin{eqnarray}
	\frac{\partial\mathcal{L}_1}{\partial\phi_1^\prime} &=& -\frac{r^4l_0^2}{R^4\sqrt{1-v^2}}[(v^2-f)\phi_1^\prime-v\xi_0^\prime]\nonumber\\
	&\equiv&\frac{C_1l_0^2}{R^4\sqrt{1-v^2}},
	\label{eqC1}
\end{eqnarray}
with $C_1$ a constant and
\begin{equation}
	\phi_1^\prime=\frac{1}{f-v^2}\left(\frac{C_1}{r^4}-v\xi_0^\prime\right).
\end{equation}
To eliminate the unphysical singularity at $f-v^2=0$, i.e., at
\begin{equation}
	r=(1-v^2)^{-\frac{1}{4}}r_t\equiv r_c,
\end{equation}
we choose $C_1=vr_c^4\xi_{0c}^\prime$ with $\xi_{0c}^\prime$ given by Eq.(\ref{eq6031}) evaluated at $r_c$. Consequently,
\begin{equation}
	\phi_1^\prime=\frac{R^2r_t^2}{r^4-r_t^4}.
\end{equation}

The second equation of Eqs.(\ref{first_integral}) with $\mathcal{L}$ replaced with $\mathcal{L}_1$ is satisfied automatically because of Eq.(\ref{eq6031}), i. e. 
\begin{equation}
	\frac{\partial\mathcal{L}_1}{\partial\xi_1^\prime}=\frac{vr_t^2}{R^2\sqrt{1-v^2}}=\rm{const.}
	\label{eq6032}
\end{equation}
The explicit forms of $\xi_1^\prime$ and $l_1^\prime$ have to be determined with the aid of $\mathcal{L}_2$ and we have
\begin{equation}
	\frac{\partial\mathcal{L}_2}{\partial\xi_1^\prime}=\frac{r^4(f-v^2)}{R^4(1-v^2)^{\frac{3}{2}}}\xi_1^\prime+\frac{vl_0^2r_t^2}{2R^2(1-v^2)^{\frac{3}{2}}}\equiv C_2,
	\label{eqC2}
\end{equation}
with $C_2$ a constant. Note that the second term on LHS is already a constant. To eliminate the singularity at $f-v^2=0$ in the solution for $\xi_1^\prime$, $C_2$ has to cancel it and we end up with
\begin{equation}
	\xi_1^\prime=0.
\end{equation}

Coming to the determination of $l_1^\prime$, we have
\begin{equation}
	\frac{\partial\mathcal{L}_2}{\partial l_1^\prime}=\frac{r^4(f-v^2)}{R^4\sqrt{1-v^2}}l_1^\prime,
\end{equation}
\begin{equation}
	\frac{\partial\mathcal{L}_2}{\partial l_1}=-\frac{l_0}{\sqrt{1-v^2}},
\end{equation}
and Eq.(\ref{EL4}) becomes
\begin{equation}
	\label{eq609}
	l_{1}^{\prime\prime}+\frac{4(1-v^{2})}{r(f-v^{2})}l_{1}^{\prime}+\frac{R^{4}l_{0}}{r^{4}(f-v^{2})}=0,
\end{equation}
the solution to the above equation is
\begin{equation}
	\label{eq610}
	l_{1}^\prime=\frac{R^4}{r^4}\frac{r_c-r}{f-v^2}l_0=-\frac{R^4}{(1-v^2)(r+r_c)(r^2+r_c^2)}l_0,
\end{equation}
where we used the same method as above to determine the constant of integration.

Substituting Eqs.(\ref{eqC1}), (\ref{eq6032}), (\ref{eqC2}) and (\ref{eq610}) together with the explicit expressions of $\xi_1$, $C_1$ and $C_2$ into Eqs.(\ref{eq6022}), we end up with different components of the drag force 
\begin{equation}
	\begin{split}
		\label{eq611}
		&\mathfrak{f}_z=-\frac{1}{2\pi\alpha^{\prime}}\frac{r_{t}^{2}}{R^{2}}\frac{v}{\sqrt{1-v^{2}}}\left(1+\frac{\omega^{2}l_{0}^{2}}{2}\frac{1}{1-v^{2}}\right),
		\\&\mathfrak{f}_\phi =-\frac{1}{2\pi\alpha^{\prime}}\omega l_{0}^{2}\frac{r_{t}^{2}}{R^{2}}\frac{1}{\sqrt{1-v^{2}}},	\\&\mathfrak{f}_l=-\frac{1}{2\pi\alpha^{\prime}}\omega^{2}l_0\frac{1}{\sqrt{1-v^{2}}}\left[\frac{r_t}{(1-v^2)^{\frac{1}{4}}}-r\right].
	\end{split}
\end{equation}
Similar to the drag force calculated in Kerr-$\text{AdS}_5$ \cite{Arefeva:2020jvo}, there is a linearly divergent term in the $l$ component of Eq.(\ref{eq611}) with $r\rightarrow \infty$. One can renormalize it by introducing a large but finite mass $m_{rest}$ of the heavy quark located at $r_m=2\pi\alpha^{\prime} m_{rest}$. Using Eq.(\ref{NambuGoto}) and Eq.(\ref{lambda}), the final result for the drag force is
\begin{equation}
	\begin{split}
		\label{eq612}
		&\mathfrak{f}_z=-\frac{\pi \sqrt{\lambda}T^2}{2}\frac{v}{\sqrt{1-v^{2}}}\left(1+\frac{\omega^{2}l_{0}^{2}}{2}\frac{1}{1-v^{2}}\right),
		\\&\mathfrak{f}_\phi =-\frac{\pi \sqrt{\lambda}T^2}{2}\omega l_{0}^{2}\frac{1}{\sqrt{1-v^{2}}},
		\\&\mathfrak{f}_l=\omega^{2}l_0\frac{1}{\sqrt{1-v^{2}}}\left[m_{rest}-\frac{T\sqrt{\lambda}}{2(1-v^2)^{\frac{1}{4}}}\right].
	\end{split}
\end{equation}

\begin{figure}
	\centering
	\includegraphics[scale=1]{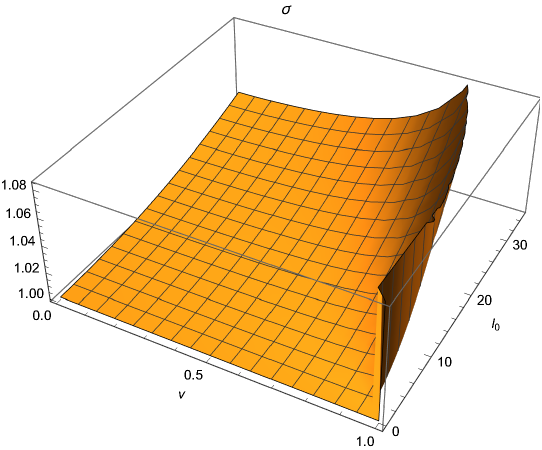}
	\caption{The ratio $\sigma=\mathfrak{f}_z/\mathfrak{f}_{\omega=0}$ as a function of the velocity $v$ and the radius $l_0$, here we set $\omega=7$ MeV.}
	\label{fig11}
\end{figure}

Setting $\omega=0$, we recover the drag force calculated in \cite{Gubser:2006bz} and \cite{Herzog:2006gh} as expected. For $\omega\ne 0$, the correction term of $\mathfrak{f}_z$ suggests an enhanced dissipation mechanism. The physical meaning of various components of (\ref{eq612}) is best understood for a non-relativistic motion of the heavy quark, i.e. $v\ll1$, where Eq.(\ref{eq612}) becomes
\begin{equation}
	\begin{split}
		&\mathfrak{f}_z=-\frac{\pi \sqrt{\lambda}T^2}{2}v\left(1+\frac{\omega^{2}l_{0}^{2}}{2}\right),
		\\&\mathfrak{f}_\phi =-\frac{\pi \sqrt{\lambda}T^2}{2}\omega l_{0}^{2},
		\\&\mathfrak{f}_l=\omega^{2}l_0\left(m_{rest}-\frac{T\sqrt{\lambda}}{2}\right).
	\end{split}
	\label{eq614}
\end{equation}
The component $\mathfrak{f}_\phi/l_0$ together with the first term of $\mathfrak{f}_z$ of Eq.(\ref{eq614}) are simply the transformation of the drag in the static frame via the non-relativistic transformation of the velocity, i.e. $\vec v\to\vec v+\omega\hat z\times\vec l_0$. The second term of $\mathfrak{f}_z$ represents additional dissipation by a rotating medium. The component $\mathfrak{f}_l$ is nothing but the centrifugal force subject to a mass renormalization by the medium. The other inertial force, the Coriolis force, is absent since the velocity of the heavy quark in the rotating frame is parallel to the rotation axis.

Numerically, the angular velocity $\omega$ estimated in \cite{STAR:2017ckg} is around 7 MeV. Taking the radius transverse to the rotation axis to be 7 fm (the radius of a gold nucleus), we have $\omega l_0<0.25$ and the small $\omega$ appears not too crude. The ratio of $\mathfrak{f}_z$ with rotation to the drag force without rotation in terms of $v$ and $l_0$ is plotted in Fig.\ref{fig11}.

According to Eq.(\ref{eq614}), the friction coefficient defined in \cite{Herzog:2006gh} reads
\begin{equation}
	\mu =-\frac{\mathfrak{f}_z}{p}=\frac{\pi \sqrt{\lambda}T^2}{2m}\left(1+\frac{\omega^{2}l_{0}^{2}}{2}\right), 
\end{equation}
with $m$ the heavy quark mass. If we set the temperature $T=0.2$ GeV, the linear velocity $\omega l_0=0.2$, and the 't Hooft coupling constant $5.5<\lambda <6\pi$ \cite{Gubser:2006qh}. We find 0.118 GeV$<\mu <$0.218 GeV for $c$ quark and 0.036 GeV$<\mu <$0.067 GeV for $b$ quark, where the masses of $c$ quark and $b$ quark are $m_c=1.275$ GeV and $m_b=4.18$ GeV respectively \cite{ParticleDataGroup:2018ovx}.

What we worked out here is the simplest setup for the drag force, where the distance of the moving heavy quark to the rotation axis ($z$-axis), $l$, is a constant in time. This is not the case if the heavy quark moves in other directions. Consider the case when the heavy quark moves along the radial direction, $ l=vt$ on the boundary. The dependence of the background metric (\ref{eq222}) on $l$ would render the world-sheet metric explicitly time dependent. Consequently, the Euler-Lagrange equation would be a set of genuine partial differential equations with respect to $r$ and $t$. This would add difficulties to the solution, though still tractable under small $\omega$ approximation. Our progress for a general setup of the drag force, if any, will be reported separately.

\section{Heavy quark potential in global rotating background}\label{sec:04}

In this section, we explore the impact of rotation on the heavy quark potential in the global rotation background. Because of the inhomogeneity and anisotropy introduced by the rotation, the energy of a heavy quarkonium depends not only on the distance between the constituent quark and antiquark but also on its location (e.g., the center of mass) and orientation (the direction of the link between the quark and antiquark). For a given location and orientation, we define the heavy quark potential as the difference between the energy of the quarkonium and the self-energies of the constituent quark and antiquark. Both the energy of the quarkonium and the self-energy of a single quark can be extracted from the expectation value of the Wilson loop/line operators as 
\begin{equation}
	\label{eq32}
	\langle W(C)\rangle= e^{-i\mathcal{T}E[C]},
\end{equation}
with $\mathcal{T}\to\infty$ the time extension. For a quarkonium, $C$ consists of a pair of straight lines parallel to the time axis and separated by the distance between the quark and the antiquark. For a single quark or antiquark, $C$ is a straight line parallel to the time axis. The heavy quark potential is defined as $V(L)=E[C({\rm{quarkonium}})]-2E[C(\rm{single})]$ with $L$ the interquark distance.

\begin{figure}
	\centering
	\includegraphics[scale=0.6]{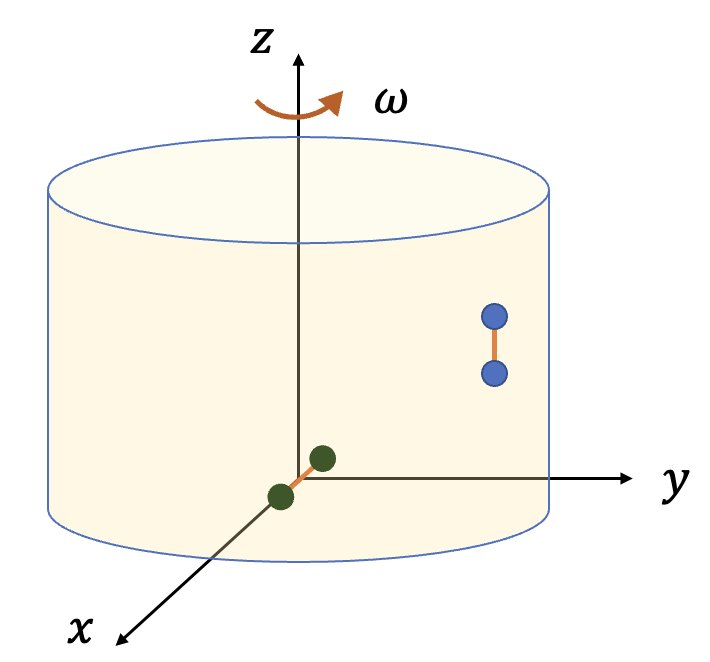}
	\caption{A sketch of the position of the quarkonium in QGP.}
	\label{fig1}
\end{figure}

Holographically, the Wilson-loop expectation value of the super Yang-Mills in the large $N_{c}$ and large $\lambda$ limit corresponds to the on-shell Nambu-Goto action (i.e., the action evaluated at the solution of the Euler-Lagrange equations) of the world-sheet embedded in the 5D background and fixed on the $\text{AdS}_5$ boundary to the contour $C$, i.e.  
\begin{equation}
	\label{eq33}
	\langle W(C)\rangle\sim e^{iS[C]}.
\end{equation}
The shape of the world-sheet is the solution of the Euler-Lagrange equation. The world-sheet corresponding to a quarkonium is a U-shaped cylinder outside the black hole horizon, while that corresponding to a single (anti-)quark extends from the boundary to the horizon. In what follows, we shall denote the Nambu-Goto action of the quarkonium by $S_U$ and that of a single quark and a single antiquark by $S_{\parallel}$ with the subscript signifies two parallel world-sheets, one for a single quark and one for a single antiquark, which is simply twice of the Nambu-Goto action of a single quark.  Consequently
\begin{equation}
	V(L)=-\frac{1}{\mathcal{T}}\left(S_U-S_\parallel\right).
	\label{potential}
\end{equation}
Even though the rotation introduces the coupling among different components of the Euler-Lagrange equation, the analysis in Appendix A shows that the zeroth order world-sheet suffices to determine $S_U$ up to the order $\omega^2$. This is, however, not the case with the world-sheet of a single quark because of an infrared singularity at the horizon. In what follows, we shall work out integral representation of the heavy quark potential up to the order of $\omega^2$ for two special cases shown in Fig.\ref{fig1}, where the quarkonium is parallel to the axis of rotation or is symmetric with respect to it. The heavy quark potential associated to the quarkonium at an arbitrary location with an arbitrary orientation is discussed in Appendix B. 

\subsection{Parallel to the rotation axis}\label{sec:4.1}

In the absence of rotation, the background metric is homogeneous and isotropic with respect to the boundary coordinates and one may embed the world-sheet corresponding to a quarkonium in the background metric as $\phi=\rm{const.}$, $l=l_0=\rm{const.}$, and $z=z(r)$, retaining $t$ and $r$ as the world-sheet coordinates. Substituting into the background metric Eq.(\ref{eq22}), we find that
\begin{equation}
	-g=\frac{r^4}{R^4}\left(f z^{\prime 2}+\frac{R^4}{r^4}\right).
\end{equation}
The Euler-Lagrange equation of the Nambu-Goto action yields a U-shaped profile
\begin{equation}
	z=\pm R^2\int_{r_c}^r d\mathbf{r} \sqrt{\frac{h(r_c)}{h(\mathbf{r})(h(\mathbf{r})-h(r_c))}},
	\label{U_profile}
\end{equation}
where 
\begin{equation}
	h(r)=r^4f(r),
	\label{h(r)}
\end{equation}
for the two quarks located at $z=\pm(L/2)$ on the boundary. We have 
\begin{equation}
	L=2R^2\int_{r_c}^\infty dr\sqrt{\frac{h_c}{h(h-h_c)}}=\frac{\mathcal{G}(b)}{T},
	\label{LT_b}
\end{equation}
which determines $b\equiv r_t/r_c<1$ as an implicit function of $LT$ with
\begin{equation}
	\mathcal{G}(b)\equiv\frac{2b\sqrt{1-b^4}}{\pi}\int_1^\infty\frac{d\eta}{\sqrt{(\eta^4-1)(\eta^4-b^4)}},
	\label{functionG}
\end{equation}
and we have used $h=h(r), h_c=h(r_c)$ for convenience. The zeroth order Nambu-Goto action
\begin{equation}
	S_U^{(0)}=-\frac{\mathcal{T}}{\pi\alpha^\prime}\int_{r_c}^K dr \sqrt{\frac{h}{h-h_c}},
\end{equation}
here the UV cutoff $K\rightarrow \infty $ in the end. Switching on the rotation and maintaining the same embedding, the metric determinant becomes
\begin{equation}
	-g=\frac{r^4}{R^4}(f-\omega^2l_0^2)\left(z^{\prime 2}+\frac{R^4}{fr^4}\right),
\end{equation}
and the corresponding Nambu-Goto action becomes
\begin{equation}
	S_U=-\frac{\mathcal{T}}{\pi\alpha^\prime}\int_{r_c}^K dr\left(\sqrt{\frac{h}{h-h_c}}-\frac{1}{2}\omega^2 l_0^2\frac{1}{f}\sqrt{\frac{h}{h-h_c}}+O(\omega^4)\right),
	\label{pair}
\end{equation}
The corrections to the zeroth order solution contribute $O(\omega^4)$ terms of $S_U$.

Unlike $S_U$, the zeroth order world-sheet for a single (anti-)quark with constant $z$, $\phi$ and $l$ extending to the black hole horizon $r_t$ cannot be employed to calculate $S_\parallel$ to the order $\omega^2$ because of the singularity at $r_t$. Instead, we explore the alternative world-sheet obtained by taking the limit $v\to 0$ of the drag force solution \footnote{See Ref.\cite{Avramis:2006em} for a similar solution in the context of nonzero R-charge.}. We have
\begin{equation}
	\phi^\prime = \omega\frac{R^2r_t^2}{r^4f}, \qquad z^\prime = 0, \qquad l^\prime = -\omega^2\frac{R^4(r-r_t)}{r^4f}l_0.
\end{equation}
The contribution of $l^\prime$ to the Nambu-Goto action is of the order $\omega^4$ and thereby can be ignored here. After miraculous cancellations, we end up with
\begin{equation}
	-g=1-\omega^2l_0^2,
	\label{single}
\end{equation}
for the induced metric of the revised single quark world-sheet and then
\begin{equation}
	S_\parallel=-\frac{\mathcal{T}}{\pi\alpha^\prime}\int_{r_t}^K dr \left(1-\frac{1}{2}\omega^2l_0^2+O(\omega^4)\right).
\end{equation} 

\begin{figure}
	\centering
	\includegraphics[scale=0.9]{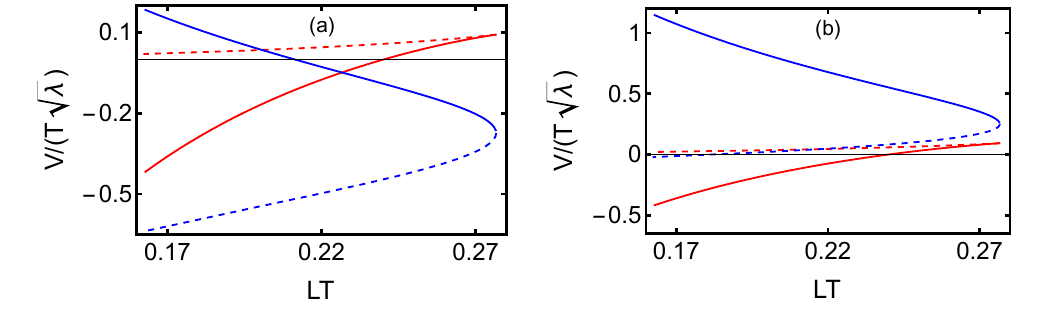}
	\caption{The heavy quark potential $V$ as function of the distance between the quark and antiquark $L$. The red line represents $V_0/(T\sqrt{\lambda})$, the blue line represents $V_1/(T\sqrt{\lambda})$. The solid line represents the physical branch and the dotted line represents the unphysical branch. (a) the quarkonium is parallel to the rotation axis, (b) symmetric with respect to the rotation axis.}
	\label{fig2}
\end{figure}

It follows from Eq.(\ref{potential}) that the heavy quark potential in this case
\begin{equation}
	\label{V_global}
	V(L)=V_0(L)+\omega^2l_0^2V_1(L)+O(\omega^4),
\end{equation}
where the potential without rotation \cite{Rey:1998bq,Liu:2006he,Avramis:2006em,Brandhuber:1998bs}  
\begin{equation}
	V_0(L)
	=\frac{1}{\pi\alpha^\prime}\int_{r_c}^\infty dr\left(\sqrt{\frac{h}{h-h_c}}-1\right)-\frac{r_c-r_t}{\pi\alpha^\prime}=\sqrt{\lambda}T\mathcal{F}(b),
	\label{V_0z}
\end{equation}
with
\begin{equation}
	\mathcal{F}(b)\equiv \frac{1}{b}\left[\int_1^\infty d\eta\left(\sqrt{\frac{\eta^4-b^4}{\eta^4-1}}-1\right)-1+b\right],
	\label{functionF}
\end{equation} 
and the coefficient of the leading order correction by rotation  
\begin{equation}
	\begin{split}
		V_1(L)
		=-\frac{1}{2\pi\alpha^\prime}\int_{r_c}^\infty dr\left(\frac{1}{f}\sqrt{\frac{h}{h-h_c}}-1\right)+\frac{r_c-r_t}{2\pi\alpha^\prime}
		=-\sqrt{\lambda}T\mathcal{P}(b),
		\label{V_1z}
	\end{split}
\end{equation}
with
\begin{equation}
	\mathcal{P}(b)\equiv\frac{1}{2b}\left[\int_1^\infty d\eta\left(\frac{\eta^4}{\sqrt{(\eta^4-1)(\eta^4-b^4)}}-1\right)-1+b\right].
\end{equation}  
In the UV limit where $L\to 0$, $r_c\to\infty$ and $V_0(L)$ approaches to the heavy quark potential at zero temperature \cite{Maldacena:1998im}. Furthermore in this limit, $f\simeq 1$ throughout the integration domain of Eq.(\ref{V_1z}), so $V_1(L)\simeq -\frac{1}{2}V_0(L)$. The asymptotic form of the heavy quark potential is then
\begin{equation}
	V(L)\simeq-\frac{4\pi^2\sqrt{\lambda}}{\Gamma^4\left(\frac{1}{4}\right)L}\left(1-\frac{1}{2}\omega^2l_0^2+O(\omega^4)\right),
\end{equation}
and the rotation effect reduces the depth of the potential well. On the other hand, at the distance $L=L_0$ where $V_0(L_0)=0$, we have
\begin{equation}
	V_1(L_0)=-\frac{1}{2\pi\alpha^\prime}\int_{r_c}^\infty dr\left(\frac{1}{f}-1\right)\sqrt{\frac{h}{h-h_c}}<0.
\end{equation}
If we define the range of the potential by the condition that $V(L)=0$, the rotation increases the range by
\begin{equation}
	\delta L=\frac{|V_1(L_0)|}{V_0^\prime(L_0)}\omega^2l_0^2.
\end{equation}
As shown in Fig.\ref{fig2} (a), the correction term $V_1>0$ when $L$ is small and $V_1<0$ when $L$ is large, consistent with the above analysis. This is different from all other holographic rotation formulations where the correction of the heavy quark potential by rotation is always positive. As estimated at the end of section \ref{sec:03}, for $\omega\simeq 7$ MeV and $l_0<7$ fm, the small $\omega$ approximation appears reasonable.

\subsection{Symmetric with respect to the rotation axis} 
\label{sec:4.2} 

For the special case where the two quarks are located at $l=L/2$ and $\phi=0,\pi$, it is more convenient to work with the metric with Cartesian coordinates on the boundary, i.e. the second and third line of Eq.(\ref{eq222}).
The world-sheet of a quarkonium at $\omega=0$ corresponds to the embedding $y=0$, $z=\rm{const.}$ and $x=x(r)$ with $|x|\le L/2$. The Euler-Lagrange equation yields the zeroth order solution which is simply Eq.(\ref{U_profile}) with $z$ replaced by $x$, i.e.
\begin{equation}
	x=\pm R^2\int_{r_c}^r d\mathbf{r} \sqrt{\frac{h_c}{h(\mathbf{r})(h(\mathbf{r})-h_c)}}\equiv\pm\frac{L}{2}u\left(\frac{r}{r_c}\right),
\end{equation}
where
\begin{equation}
	u(\eta)=\frac{2b\sqrt{1-b^4}}{\pi TL}\int_1^\eta\frac{dy}{\sqrt{(y^4-1)(y^4-b^4)}},
\end{equation}
and $0<u(\eta)<1$. Maintaining the same embedding at $\omega\ne 0$, the determinant of the induced metric becomes
\begin{equation}
	g=-\frac{r^4}{R^4}(f-\omega^2x^2){\left(x^{\prime 2}+\frac{R^4}{fr^4}\right)},
\end{equation}
and the Nambu-Goto action for the U-shaped world-sheet reads
\begin{equation}
	S_U=-\frac{\mathcal{T}}{\pi\alpha^\prime}\int_{r_c}^K dr\left(\sqrt{\frac{h}{h-h_c}}-\frac{1}{2}\omega^2\frac{x^2}{f}\sqrt{\frac{h}{h-h_c}}+O(\omega^4)\right).
\end{equation}
The single quark contribution remains unchanged as it depends only on the distance to the rotation axis. The determinant of the induced metric of the single quark world-sheet is thereby given by Eq.(\ref{single}) with $l_0=L/2$ and the Nambu-Goto action pertaining to the single quark and antiquark is
\begin{equation}
	S_\parallel=-\frac{\mathcal{T}}{\pi\alpha^\prime}\int_{r_t}^K dr \left(1-\frac{1}{8}\omega^2L^2+O(\omega^4)\right).
\end{equation}

Following Eq.(\ref{potential}), the heavy quark potential in this case reads
\begin{equation}
	V(L)=V_0(L)+\frac{1}{4}\omega^2L^2V_1(L)+O(\omega^4),
\end{equation}
while the heavy quark potential of zeroth $V_0(L)$ is intact, the coefficient of the rotation correction becomes
\begin{eqnarray}
	V_1(L)&=&-\frac{1}{2\pi\alpha^\prime}\int_{r_c}^\infty dr\left(\frac{4x^2}{L^2f}\sqrt{\frac{h}{h-h_c}}-1\right)+\frac{r_c-r_t}{2\pi\alpha^\prime} \nonumber  \\
	&=& -\frac{\sqrt{\lambda}T}{2b}\left[\int_1^\infty d\eta\left(\frac{u^2(\eta)\eta^4}{\sqrt{(\eta^4-1)(\eta^4-b^4)}}-1\right)-1+b\right].
	\label{V_1_radial}
\end{eqnarray}
In the UV limit, $L\to 0$ and $b\to 0$ under the integration. It follows that
\begin{equation}
	V_1(L)>-\frac{1}{2}V_0(L),
\end{equation} 
which implies elevation of the potential well. Transforming the integration variable according to $r=r_c\sec\theta$ and utilizing the relationship Eq.(\ref{LT_b}) at $f=1$, we have
\begin{equation}
	V_1(L)\simeq \frac{\sqrt{2\pi\lambda}\kappa}{\Gamma^2\left(\frac{1}{4}\right)L},
\end{equation}
where
\begin{equation}
	\kappa =1-\int_0^{\frac{\pi}{2}}d\theta \sec^2 \theta \left\{\frac{\Gamma^4 \left(\frac{1}{4}\right) \left[2E(\theta,1/2)-F(\theta,1/2)\right]^2}{4\pi^3\sqrt{1+\cos^2\theta}}-\sin\theta\right\} \simeq 1.42,
\end{equation}
with $F(\theta,m)$ and $E(\theta,m)$ standing for the incomplete elliptic integrals of the first kind and the second kind.
The asymptotic behavior of the heavy quark potential in this case reads
\begin{equation}
	V(L)\simeq-\frac{4\pi^2\sqrt{\lambda}}{\Gamma^4\left(\frac{1}{4}\right)L}+\frac{\sqrt{2\pi\lambda}\kappa}{4\Gamma^2\left(\frac{1}{4}\right)}\omega^2L.
\end{equation}
On the IR side, at the critical length $L_0$ when $V_0(L_0)=0$, numerical integration of Eq.(\ref{V_1_radial}) shows that $V_1(L_0)>0$, implying a shortening of the potential range by
\begin{equation}
	\delta L=-\frac{V_1(L_0)}{V_0^\prime(L_0)}\omega^2l^2.
\end{equation}
Fig.\ref{fig2} (b) shows that the correction term $V_1>0$, the heavy quark potential in this case is entirely above that in the absence of rotation. Thereby, the potential well becomes shallower because of rotation. For $\omega\simeq 7$ MeV and the temperature $T=0.2$ GeV, the expansion parameter $\omega L\simeq 0.01$ is indeed very small, justifying the small $\omega$ approximation.

\section{Other holographic formulation of rotation}\label{sec:05}

In the last two sections, we explored a rotating QGP in terms of the metric Eq.(\ref{eq222}) and calculated the drag force and the heavy quark potential to the leading order correction of the angular velocity. As was argued in Section \ref{sec2}, such a holographic formulation is inspired by an ensemble of thermal equilibrium with a macroscopic angular momentum with respect to the laboratory frame. In this section, we shall comment on the other two holographic rotation formulations discussed in the literature. One is the local Lorentz transformation and the other is Kerr-$\text{AdS}_5$.

\subsection{Local Lorentz transformation} \label{sec:051}	

The holographic formulation proposed in \cite{Trocheris1949CIV,0On} and employed in \cite{BravoGaete:2017dso,Erices:2017izj,Awad:2002cz,Zhao:2022uxc,Chen:2020ath,Chen:2022mhf,Zhou:2021sdy,Zhang:2023psy} amounts to the following  local Lorentz transformation on the static metric Eq.(\ref{eq22})
\begin{equation}
	\label{eq23}
	t\longrightarrow\frac{1}{\sqrt{1-(\omega l_{0})^{2}}}(t+\omega l_{0}^{2} \phi),\quad \phi\longrightarrow\frac{1}{\sqrt{1-(\omega l_{0})^{2}}}(\phi+\omega t),
\end{equation}
where $\omega$ is the angular velocity, and $l_{0}$ is a constant length. The resulting metric is
\begin{eqnarray}
	\label{eq24}
	ds^{2}&=&\frac{r^{2}}{R^{2}}\frac{1}{1-(\omega l_{0})^{2}}[(\omega^{2}l^{2}-f(r))dt^{2}
	+(l^{2}-\omega^{2}l_{0}^{4}f(r))d\phi^{2}
	+2\omega(l^{2}-l_{0}^{2}f(r))dtd\phi]\nonumber\\&+&\frac{r^{2}}{R^{2}}(dl^{2}+dz^{2})+\frac{1}{f(r)}\frac{R^{2}}{r^{2}}dr^{2}.
\end{eqnarray}
Using the formula in \cite{ZHAO1983201}, the Hawking temperature of the black hole is given by
\begin{equation}
	\label{eq27}
	T=|\frac{\lim\limits_{r\to r_{t}}-\frac{1}{2}\sqrt{\frac{g^{rr}}{-\hat{g}_{00}}}\hat{g}_{00,1}}{2\pi}|=\frac{r_{t}\sqrt{1-(\omega l_0)^{2}}}{\pi R^{2}}.
\end{equation}

As the speed $\omega l_0$ in the transformation Eq.(\ref{eq23}) only matches the linear speed of rotation at a particular radial coordinate $l=l_0$ and Eq.(\ref{eq23}) does not map the full angle range $\Delta\phi=2\pi$ before the transformation to the full angle range $\Delta\phi=2\pi$ afterwards, the metric can only describe a small neighborhood around $l=l_0$ and a domain of $\phi$ less than $2\pi$ wide in the rotating QGP. Because of this limitation, the formulation can only be employed to calculate the heavy quark potential with the link between the quark and antiquark parallel to the rotation axis at a distance $l_0$ away from it.  

\begin{figure}
	\centering
	\includegraphics[scale=0.45]{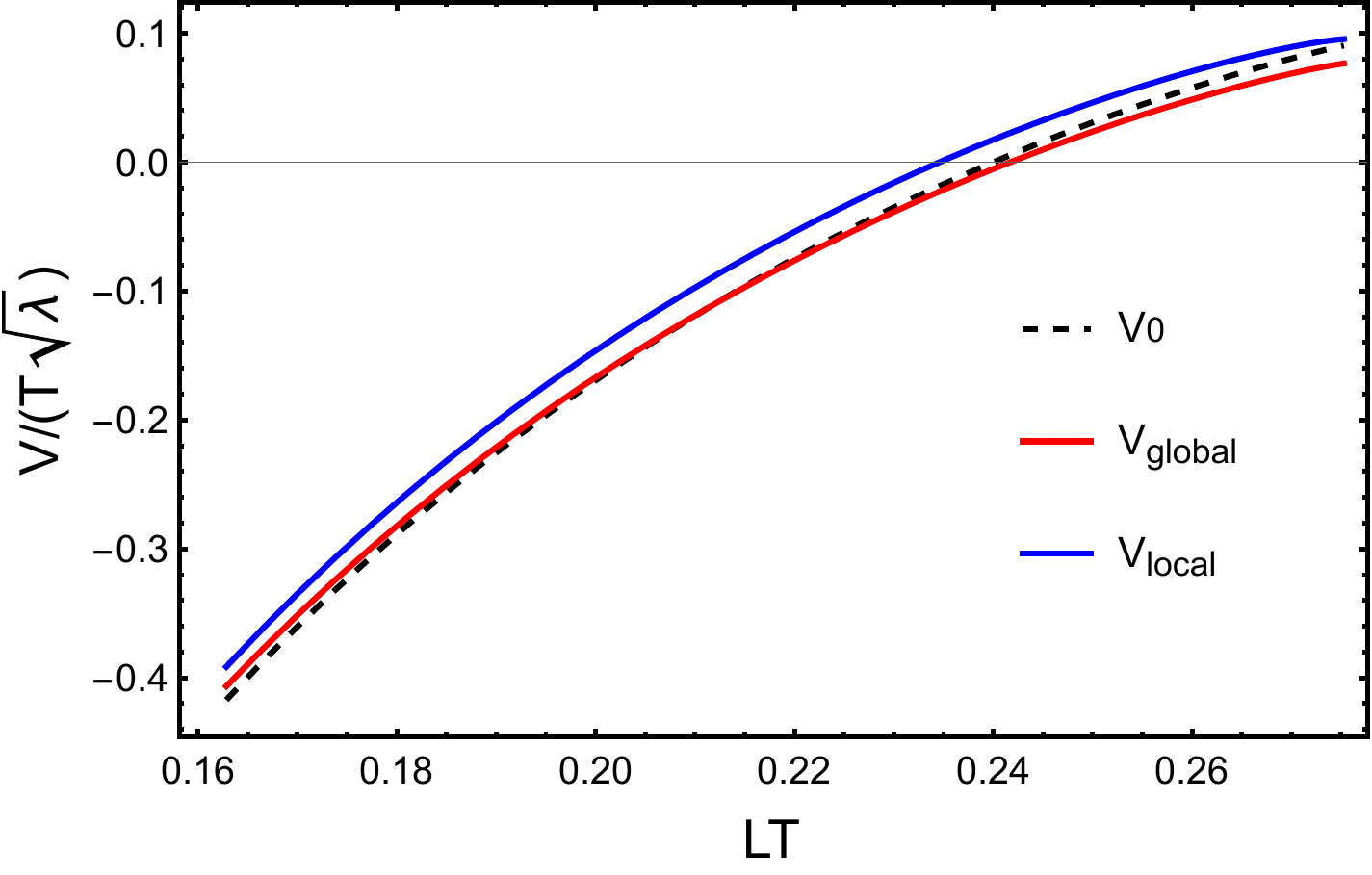}
	\caption{The heavy quark potential $V$ in different formulations versus the distance between the quark and antiquark $L$ for $\omega l_0=0.24$, where the heavy quark potential in the absence of rotation $V_0$ is also displayed for reference.}
	\label{fig4}
\end{figure}

Similar to the global rotation background, the solution U-shaped world-sheet of the form $z=z(r)$, $\phi=\phi_0=\rm{const.}$ and $l=l_0=\rm{const.}$ no longer solves the Euler-Lagrange equation for nonzero $\omega$ and different components of the equation couple with each other. The solution of the equations of motion becomes complex because of rotation. For small $\omega$ approximation, however, the $O(\omega^2)$ correction can be obtained by substituting the U-shaped world-sheet at $\omega=0$, Eq.(\ref{U_profile}) together with $l=l_0=\rm{const.}$ into the metric Eq.(\ref{eq24}) and expanding the resulting Nambu-Goto action up to $O(\omega^2)$. We have
\begin{eqnarray}
	S_U&=&-\frac{\mathcal{T}}{\pi\alpha^\prime}\int_{r_c}^K dr\sqrt{\frac{\left(1-\frac{\omega^2l_0^2}{f}\right)h}{(1-\omega^2l_0^2)(h-h_c)}}\nonumber\\
	&=&-\frac{\mathcal{T}}{\pi\alpha^\prime}\int_{r_c}^K dr\left[\sqrt{\frac{h}{h-h_c}}+\frac{1}{2}\omega^2l_0^2\left(1-\frac{1}{f}\right)\sqrt{\frac{h}{h-h_c}}\right],
\end{eqnarray}
together with the world-sheet solution of a single quark, $l=l_0$, $z=\pm L/2$ and
\begin{equation}
	\phi^\prime=\frac{\omega}{\sqrt{1-\omega^2l_0^2}}\frac{R^2r_t^2}{r^4f},
\end{equation}
which implied that $\sqrt{-g}=1$. It follows that the heavy quark potential in the local Lorentz frame is
\begin{eqnarray}
	V_{\rm{local}}(L)&=&\frac{1}{\pi\alpha^\prime}\int_{r_c}^\infty dr\left[\left(\sqrt{\frac{h}{h-h_c}}-1\right)+\frac{1}{2}\omega^2l_0^2\left(1-\frac{1}{f}\right)\sqrt{\frac{h}{h-h_c}}\right]-\frac{r_c-r_t}{\pi\alpha^\prime}\nonumber\\
	&=&\frac{\sqrt{\lambda} T}{\sqrt{1-\omega^2l_0^2}}[\mathcal{F}(\beta)+\omega^2l_0^2\mathcal{Q}(\beta)+...],
	\label{V_local}
\end{eqnarray}
where $\beta=r_t/r_c$, the function $\mathcal{F}$ is defined in (\ref{functionF}), and  
\begin{equation}
	\mathcal{Q}(\beta)=-\frac{\beta^3}{2}\int_1^\infty\frac{d\eta}{\sqrt{(\eta^4-1)(\eta^4-\beta^4)}}.
\end{equation}
Here we have used Eq.(\ref{lambda}) and Eq.(\ref{eq27}). The ignored terms inside the bracket of (\ref{V_local}) are higher order terms beyond $\omega^2$. Because of different relations between the temperature and the horizon radius in Eq.(\ref{eq223}) and Eq.(\ref{eq27}), we use $\beta$ here to distinguish from $b$ in Section \ref{sec:04}. For the same reason, the equation (\ref{LT_b}) becomes
\begin{equation}
	L=\frac{\sqrt{1-\omega^2l_0^2}}{T}\mathcal{G}(\beta),
\end{equation}
which determines $\beta$ in terms of $LT$. We have 
\begin{equation}
	\beta=b+\frac{LT}{2\mathcal{G}^\prime(b)}\omega^2l_0^2,
\end{equation}
to the order of $\omega^2$. Substituting into (\ref{V_local}) and keeping only the terms up to $\omega^2$, we find that
\begin{equation}
	V_{\rm{local}}(L)=\sqrt{\lambda}T\left\{\mathcal{F}(b)+\omega^2l_0^2\left[\frac{1}{2}\mathcal{F}(b)+\frac{\mathcal{G}(b)\mathcal{F}^\prime(b)}{2\mathcal{G}^\prime(b)}-\mathcal{Q}(b)\right]\right\},
\end{equation}
where the prime denotes the derivative with respect to $b$. This is to be compared with the heavy quark potential in the global rotation background Eq.(\ref{V_global}),
\begin{equation}
	V_{\rm{global}}(L)=\sqrt{\lambda}T\left[\mathcal{F}(b)-\omega^2l_0^2\mathcal{P}(b)\right],
\end{equation}
for the same parameters $T$, $L$ and $\lambda$. According to Fig.\ref{fig4}, the binding between quark and antiquark in $V_{\rm local}$(L) is relaxed by the rotation in contrast to $V_{\rm{global}}(L)$ \footnote{The conclusion is the same as \cite{Chen:2022obe}, but the solution ansatz in  \cite{Chen:2022obe} does not satisfy the full set of the Euler-Lagrange equation.}.

\subsection{Kerr-\texorpdfstring{$\text{AdS}_5$}{AdS5}}  \label{sec:052}

Another global holographic description of the rotating QGP is implemented via the Kerr-$\text{AdS}_5$ metric which describes a rotating black hole and its conformal boundary is $R^1\times S^3$ instead of $R^1\times R^3$. The drag force and the heavy quark potential have been calculated in this background \cite{Arefeva:2020jvo,Golubtsova:2022ldm} with QGP located on the boundary. The coupling among different components of the Euler-Lagrange equation discussed at the end of section \ref{sec:02} was taken into account and small rotation parameters were assumed in \cite{Arefeva:2020jvo}. There is only one parameter, the small angular velocity in \cite{Arefeva:2020jvo}, instead of two parameters, the small angular velocity and an arbitrary linear velocity in section \ref{sec:03} of this paper. In another word, when the rotation is turned off, the heavy quark in \cite{Arefeva:2020jvo} becomes static instead of linear motion at constant speed. For the heavy quark potential, the result in \cite{Golubtsova:2022ldm} shows that the rotation relaxes the binding within a quarkonium, in line with the result in a local Lorentz frame discussed in section \ref{sec:051} above. 

A quantitative comparison is, however, hindered for the following reasons.

1) The conformal boundary of the Kerr-$\text{AdS}_5$ metric employed in 
\cite{Arefeva:2020jvo,Golubtsova:2022ldm} is $R\times S^3$ instead of being planar.

The most general Kerr-$\text{AdS}_5$ metric, given by \cite{Hawking:1998kw}, reads
\begin{eqnarray}
	ds^2 &=& -\frac{\Delta}{\rho^2}\left(dt-\frac{a\sin^2\theta}{\Xi_a}d\phi-\frac{b\cos^2\theta}{\Xi_b}d\psi\right)^2+\frac{\Delta_\theta \sin^2\theta}{\rho^2}\left(adt-\frac{r^2+a^2}{\Xi_a}d\phi\right)^2\nonumber\\
	&+& \frac{\Delta_\theta \cos^2\theta}{\rho^2}\left(bdt-\frac{r^2+b^2}{\Xi_b}d\psi\right)^2+\frac{\rho^2}{\Delta}dr^2+\frac{\rho^2}{\Delta_{\theta}}d\theta^2\nonumber\\
	&+&\frac{1+\frac{r^2}{l^2}}{r^2\rho^2}\left[abdt-\frac{b(r^2+a^2)\sin^2\theta}{\Xi_a}d\phi-\frac{a(r^2+b^2)\cos^2\theta}{\Xi_b}d\psi\right]^2,
	\label{kerr}
\end{eqnarray}
where
\begin{equation}
	\begin{split}
		\label{delta}
		&\Delta=\frac{1}{r^2}(r^2+a^2)(r^2+b^2)\left(1+\frac{r^2}{l^2}\right)-2M,
		\\&\Delta_\theta=1-\frac{a^2}{l^2}\cos^2\theta-\frac{b^2}{l^2}\sin^2\theta,
		\\&\rho^2=r^2+a^2\cos^2\theta+b^2\sin^2\theta,
		\\&\Xi_a=1-\frac{a^2}{l^2}>0, \qquad \Xi_b=1-\frac{b^2}{l^2}>0,
	\end{split}
\end{equation}
where $M$ is the mass (scaled by the 5d gravitational constant) of the source, $l$ is the AdS radius and $(a,b)$ are two parameters characterizing rotations. The black hole horizon $r_+$ is the largest root of the equation 
\begin{equation}
	\Delta(r_+)=0,
	\label{horizon}
\end{equation}
and the Hawking temperature reads
\begin{equation}
	T=\frac{\Delta^\prime(r_+)r_+^2}{4\pi(r_+^2+a^2)(r_+^2+b^2)}.
	\label{hawking}
\end{equation}
As is well-known in literature, a planar boundary of the metric (\ref{blackbrane}) corresponds to the large black hole limit \cite{Natsuume:2014sfa,Garbiso:2020puw}, which is characterized by $r_+/l\to\infty$. The AdS radius $l$ was set to 0.55 fm in \cite{Golubtsova:2022ldm}. For a typical set of parameters they used, $T=0.17$ GeV, $a=b=0.1l$, $r_+=0.55$ fm, far from the large black hole limit. 

2) A nonzero angular velocity is not the sole reason for the coupling among different components of the Euler-Lagrange equation. The non-planar boundary geometry of $S^3$ also contributes the coupling and adds the complexity of the U-shaped string world-sheet even for the Schwarzschild-$\text{AdS}_5$ metric
\begin{equation}
	ds^2=-fdt^2+\frac{dr^2}{f}+r^2(d\theta^2+\sin^2\theta d\phi^2+\cos^2\theta d\psi^2),
	\label{SchwAdS}
\end{equation}
with
\begin{equation}
	f=1+\frac{r^2}{l^2}-\frac{2M}{r^2}.
\end{equation}
The metric (\ref{SchwAdS}) corresponds to the case $a=b=0$ of the Kerr-$\text{AdS}_5$ metric (\ref{kerr}). A generic ansatz for a static U-string is $(\tau,\sigma)=(t,\phi)$ and
\begin{equation}
	r=r(\sigma), \qquad \psi=\psi(\sigma), \qquad \theta=\theta(\sigma).
\end{equation}
The determinant of the world-sheet metric reads
\begin{equation}
	g=-fr^2(\theta^{\prime 2}+\sin^2\theta+\cos^2\theta\psi^{\prime 2})-r^{\prime 2},
\end{equation}
with the prime standing for the derivative with respect to $\sigma$.
The variation of the Nambu-Goto action yields the full set of Euler-Lagrange equation
\begin{equation}
	\begin{split}
		\frac{d}{d\sigma}\left(\frac{r^\prime}{\sqrt{-g}}\right)-\frac{\theta^{\prime 2}+\sin^2\theta+\cos^2\theta\psi^{\prime 2}}{\sqrt{-g}}\left(1+\frac{2r^2}{l^2}\right)r&=0,\\
		\frac{d}{d\sigma}\left(\frac{fr^2\cos^2\theta\psi^\prime}{\sqrt{-g}}\right)&=0,\\
		\frac{d}{d\sigma}\left(\frac{fr^2\theta^\prime}{\sqrt{-g}}\right)-\frac{fr^2(1-\psi^{\prime 2})}{2\sqrt{-g}}\sin2\theta&=0.
	\end{split}
\end{equation}
The simplistic solution ansatz employed in \cite{Golubtsova:2022ldm}, where $\theta^\prime=\psi^\prime=0$, does not satisfy the third equation except for $\theta=0,\pi/2$, and thereby the heavy quark potential there can only serve a variational ansatz even for the Schwarzschild-$\text{AdS}_5$ metric before the large black hole limit. This, however, does not exclude the possibility that the approximation is numerically sound. 

For a large but finite horizon radius, when the space curvature artifact of the boundary is subleading, the rotation contribution is also subleading and both effects may be intertwined. It would be interesting to see how to isolate the pure rotation terms from the results of \cite{Arefeva:2020jvo,Golubtsova:2022ldm,Golubtsova:2021agl} and compare with the small $\omega$ approximation in this work.

\section{Conclusion and discussion}\label{sec:06}

Let us recapitulate what we have done in this work. We start with a strongly coupled QGP carrying a macroscopic angular momentum and describe its nonperturbative dynamics holographically in terms of a black brane metric with the QGP residing on the asymptotic $\text{AdS}_5$ boundary and with the boundary coordinates in a rotating frame. Then we explored the impact of the rotation on the drag force and the heavy quark potential in terms of the Nambu-Goto action. Because of the inhomogeneity and anisotropy introduced by the rotation, the simplistic solution ansatz in the absence of rotation ceases to work and one has to solve the full set of Euler-Lagrange equations. So we did perturbatively for small angular velocity and worked out a few examples.

For a heavy quark moving in the direction of the rotation axis with a constant speed, we found that the drag force is enhanced by the rotation. The quark also experiences a centrifugal force as expected in a rotating frame.

The rotation effect on the heavy quark potential appears subtle. For a quarkonium with its quark and antiquark located symmetrically with respect to the rotation axis, the binding potential is weakened with reduced magnitude and range, in line with the results from other holographic formulatios of rotation (local Lorentz frame and Kerr-AdS.) For a quarkonium with the link between quark and antiquark aligned with the rotation axis, we found that the rotation reduces the binding force but increases the force range, different from other formulations. 

The black brane metric Eq.(\ref{blackbrane}) serves a qualitative or a semi-quantitative approximation of QCD in the plasma phase. A more realistic holographic description of QCD, AdS/QCD, requires a dilaton and other fields with the phenomenologically determined parameters. It is an important future project to apply the global rotation formulation developed in this work to AdS/QCD and to explore the rotation effect on the phase structure and other thermodynamic properties.

\appendix
\section{Small \texorpdfstring{$\omega$}{omega} expansion for U-shaped string}
\label{appendix A}

The Nambu-Goto action underlying the heavy quark potential discussed in this work is of the form
\begin{equation}
	S=\int d^2\sigma\mathcal{L}\left(\frac{\partial X}{\partial\sigma},X,\sigma|\omega\right).
\end{equation}
For small $\omega$, we have the expansion
\begin{equation}
	\mathcal{L}\left(\frac{\partial X}{\partial\sigma},X,\sigma|\omega\right)=\mathcal{L}_0\left(\frac{\partial X}{\partial\sigma},X,\sigma\right)+\omega^2\mathcal{L}_1\left(\frac{\partial X}{\partial\sigma},X,\sigma \right)+...,
\end{equation}
and correspondingly
\begin{equation}
	S=S_0+\omega^2S_1.
\end{equation}
The expansion of the solution of the Euler-Lagrange equation, a U-shaped string, reads
\begin{equation}
	X(\sigma)=X_0(\sigma)+\xi(\sigma),
\end{equation}
with $X_0(\sigma)$ the solution of the Euler-Lagrange equation at $\omega=0$ satisfying the boundary condition of $X(\sigma)$ and $\xi(\sigma)$ the perturbation satisfying homogeneous boundary conditions. We have
\begin{equation}
	\frac{\partial}{\partial\sigma^\alpha}\left(\frac{\partial\mathcal{L}_0}{\partial\frac{\partial X^\mu}{\partial\sigma^\alpha}}\right)_0-\left(\frac{\partial\mathcal{L}_0}{\partial X^\mu}\right)_0=0,
	\label{EL_0}
\end{equation}
and $\xi=0$ on the boundary, where the subscript $(...)_0$ indicates that what inside the parentheses is evaluated at $X=X_0$. 

To the order $O(\omega^2)$, the expansion of the Nambu-Goto action to the quadratic order of $\xi$ reads
\begin{eqnarray}
	S&=&S_0+\omega^2 S_1\vert_{X=X_0}
	+\int d^2\sigma \Bigg[\left(\frac {\partial \mathcal{L}_0}{\partial \frac{\partial X^\mu }{\partial \sigma^\alpha}}\right)_0\frac {\partial \xi^\mu }{\partial \sigma ^\alpha}
	+\left(\frac{\partial \mathcal{L}_0}{\partial X^\mu }\right)_0\xi^\mu \nonumber
	\\&+&\frac{1}{2}\left(\frac {\partial^2 \mathcal{L}_0}{ \partial \frac{\partial X^\mu }{\partial \sigma^\alpha}\partial \frac{\partial X^\nu }{\partial \sigma^\beta}}\right)_0\frac {\partial \xi^\mu }{\partial \sigma ^\alpha}\frac {\partial \xi^\nu }{\partial \sigma ^\beta}
	+\left(\frac {\partial^2 \mathcal{L}_0}{ \partial \frac{\partial X^\mu }{\partial \sigma^\alpha}\partial X^\nu}\right)_0\frac {\partial \xi^\mu }{\partial \sigma ^\alpha}\xi ^\nu
	+\frac{1}{2}\left(\frac {\partial^2 \mathcal{L}_0}{ \partial X^\mu \partial X^\nu}\right)_0 \xi ^\mu\xi ^\nu \nonumber
	\\&+&\omega^2 \left(\frac {\partial \mathcal{L}_1}{\partial \frac{\partial X^\mu }{\partial \sigma^\alpha}}\right)_0\frac {\partial \xi^\mu }{\partial \sigma ^\alpha}
	+\omega^2 \left(\frac{\partial \mathcal{L}_1}{\partial X^\mu }\right)_0\xi^\mu\Bigg].
\end{eqnarray}
The linear terms in the first line vanish because of the zeroth order equation Eq.(\ref{EL_0}) and the homogeneous boundary condition of $\xi$. Then the variational principle applied to the rest of the terms yields the equation of motion for $\xi$, i.e.
\begin{eqnarray}
	0&=&\frac{\partial}{\partial\sigma^\alpha}\left[\left(\frac{\partial^2\mathcal{L}_0}{\partial\frac{\partial X^\mu}{\partial\sigma^\alpha}\partial\frac{\partial X^\nu}{\partial\sigma^\beta}}\right)_0\frac{\partial\xi^\nu}{\partial\sigma^\beta}\right]
	+\frac{\partial}{\partial\sigma^\alpha}\left(\frac{\partial^2\mathcal{L}_0}{\partial\frac{\partial X^\mu}{\partial\sigma^\alpha}\partial X^\nu}\right)_0\xi^\nu
	-\left(\frac{\partial^2\mathcal{L}_0}{\partial X^\mu\partial X^\nu}\right)_0\xi^\nu
	\nonumber\\&+&\omega^2\left[\frac{\partial}{\partial\sigma^\alpha}\left(\frac{\partial\mathcal{L}_1}{\partial\frac{\partial X^\mu}{\partial\sigma^\alpha}}\right)_0-\left(\frac{\partial\mathcal{L}_1}{\partial X^\mu}\right)_0\right].
\end{eqnarray}
It follows that $\xi=O(\omega^2)$ and consequently
\begin{equation}
	S=S_0+\omega^2 S_1\Vert_{X=X_0}+O(\omega^4).
\end{equation}
Therefore the zeroth order solution $X_0$ suffices to evaluate the Nambu-Goto action to the order $\omega^2$.	

For the U-shaped string world-sheet underlying the heavy quark potential with the link between the quark and antiquark parallel to the rotation axis as in section \ref{sec:4.1}, the applicability of the above discussions is better demonstrated in terms of the world-sheet coordinates $\sigma=(t,z)$ and the background space coordinates $X=(x,y,r)$ from the metric Eq.(\ref{eq222}) with $x$, $y$ and $r$ functions of $z$ subject to the boundary condition
\begin{equation}
	r\left(\pm\frac{L}{2}\right)=K, \qquad x\left(\pm\frac{L}{2}\right)=l, \qquad y\left(\pm\frac{L}{2}\right)=0,
\end{equation}
where $K>0$ serves a UV regulator and $K\to\infty$ in the end. For the quarkonium separated perpendicular to the rotation axis as in section \ref{sec:4.2}, all we need is to switch the roles of $x$ and $z$, i.e. to take $(t,x)$ as world-sheet coordinates and $y$, $z$ and $r$ the functions of $x$. 

\section{The heavy quark potential with arbitrary location and orientation}
\label{B}

In this appendix, we calculate the potential energy of the quarkonium with arbitrary location and orientation inside a rotating QGP to the order of $\omega^2$. 

Assuming that the center of the pair is located at
\begin{equation}
	\vec X=\left(b\cos\alpha, b\sin\alpha, Z\right),
\end{equation}
with $(b,\alpha,Z)$ its cylindrical coordinates and the unit vector along the link between the quark and antiquark is given by
\begin{equation}
	\vec n=(\sin\theta\cos\phi, \sin\theta\sin\phi, \cos\theta).
\end{equation}
The coordinate of a point along the link is parametrized as
\begin{equation}
	\vec x=\vec X+\zeta\vec n, \qquad -\frac{L}{2}\le\zeta\le\frac{L}{2},
\end{equation}
and the location of the quark and antiquark corresponds to $\zeta=\pm L/2$. For the U-shaped string 
\begin{equation}
	d\vec x=\vec n d\zeta=\vec n\zeta^\prime dr.
\end{equation}
It follows from the second and third line of Eq.(\ref{eq222}) that the induced  metric reads
\begin{eqnarray}
	ds^2 &=& \frac{r^2}{R^2}\bigg\{-\left[f-\omega^2\left(b^2+2b\zeta\sin\theta\cos\beta+\zeta^2\sin^2\theta\right)\right]dt^2\nonumber\\ &+& 2\omega b\sin\theta\sin\beta\zeta^\prime dtdr+\left(\zeta^{\prime 2}+\frac{R^4}{r^4f}\right)dr^2\bigg\},
	\label{worldsheet}
\end{eqnarray}
with $\beta=\alpha-\phi$.
Substituting the zeroth order solution
\begin{equation}
	\zeta=\pm R^{2}\int_{r_{c}}^r dr^\prime \frac{1}{r^{\prime 2}}\sqrt{\frac{h(r_{c})}{f(r^\prime)[h(r^\prime)-h(r_{c})]}},
\end{equation}
with $r_c$ determined by
\begin{equation}
	L=2R^{2}\int_{r_{c}}^{\infty} dr \frac{1}{r^{2}}\sqrt{\frac{h(r_{c})}{f(r)[h(r)-h(r_{c})]}},
\end{equation}
the determinant of the metric (\ref{worldsheet}) takes the form
\begin{equation}
	g=-\frac{R^4}{r^4}\left[1-\frac{\omega^2}{f}\left(b^2+\zeta^2\sin^2\theta+2b\zeta\sin\theta\cos\beta+\frac{h_c}{h}b^2\sin^2\theta\sin^2\beta\right)\right]\frac{h}{h-h_c}.
\end{equation}
It follows that Nambu-Goto action of the U-shaped string
\begin{eqnarray}
	S_U &=& -\frac{\mathcal{T}}{2\pi\alpha^\prime}\int_{r_c}^\infty dr \left[1-\frac{\omega^2}{2f}\left(b^2+\zeta^2\sin^2\theta+2b|\zeta|\sin\theta\cos\beta+\frac{h_c}{h}b^2\sin^2\theta\sin^2\beta\right)\right]\sqrt{\frac{h}{h-h_c}}\nonumber\\
	&\quad&- \frac{\mathcal{T}}{2\pi\alpha^\prime}\int_{r_c}^\infty dr \left[1-\frac{\omega^2}{2f}\left(b^2+\zeta^2\sin^2\theta-2b|\zeta|\sin\theta\cos\beta+\frac{h_c}{h}b^2\sin^2\theta\sin^2\beta\right)\right]\sqrt{\frac{h}{h-h_c}}\nonumber
	\\&\quad&+O(\omega^4)\nonumber
	\\&=& -\frac{\mathcal{T}}{\pi\alpha^\prime}\int_{r_c}^\infty dr \left[1-\frac{\omega^2}{2f}\left(b^2+\zeta^2\sin^2\theta-\frac{h_c}{h}b^2\sin^2\theta\sin^2\beta\right)\right]\sqrt{\frac{h}{h-h_c}}+O(\omega^4), 
	\label{U_general1} 
\end{eqnarray}
where we have separated the branch of $\zeta>0$ from $\zeta<0$ but the difference is cancelled in the end.  

To remove the UV divergence pertaining to the integral of (\ref{U_general1}), we need to subtract the action of the world-sheet corresponding to statics quarks at $\zeta=\pm\frac{L}{2}$. 
Following discussion in section \ref{sec:4.1}, $\sqrt{-g}$ of each quark is given by 
\begin{equation}
	\sqrt{-g}=1-\frac{\omega^2}{2}l_\pm^2+O(\omega^4),
\end{equation}
with
\begin{equation}
	l_\pm^2=b^2+\frac{1}{4}L^2\sin^2\theta\pm bL\sin\theta\cos\beta,
\end{equation}
the distance to the rotation axis. The $\pm$ term above will be cancelled between the quark and antiquark. Consequently, the heavy quark potential inside a quarkonium at arbitrary location and with arbitrary orientation is thereby given by
\begin{eqnarray}
	V(L) &=& \frac{1}{\pi\alpha^\prime}\int_{r_c}^\infty dr \bigg\{\left[1-\frac{\omega^2}{2f}\left(b^2+\zeta^2\sin^2\theta-\frac{h_c}{h}b^2\sin^2\theta\sin^2\beta\right)\right]\sqrt{\frac{h}{h-h_c}}\nonumber\\
	&-& \left[1-\frac{\omega^{2}}{2}\left(b^2+\frac{1}{4}L^2\sin^2\theta\right)\right]\bigg\}-\left[1-\frac{\omega^{2}}{2}\left(b^2+\frac{1}{4}L^2\sin^2\theta\right)\right]\frac{r_c-r_t}{\pi\alpha^\prime} 
	\label{U_general2}.
\end{eqnarray}

If we set $\theta=0,b=l_0$, the quarkonium is parallel to the rotation axis, which is the same as section \ref{sec:4.1}, and the heavy quark potential Eq.(\ref{U_general2}) reduces to the same expression there. If we set $\theta=\frac{\pi}{2},b=0$, the quarkonium is perpendicular to the rotation axis, which is the same as section \ref{sec:4.2} and gives the same result.

\section*{Acknowledgements}

We thank Xin-Li Sheng and Yan-Qing Zhao  for useful  discussions. An enlightening email communication from M. Chernodub is warmly acknowledged. This work is supported by the National Key Research and Development Program of China (No. 2022YFA1604900). This work also  is supported by the National Natural Science Foundation of China (NSFC) under Grant Nos. 12275104, 11890711, 11890710.

	
	
	
\bibliographystyle{utphys}
\bibliography{ref}

\end{document}